\documentclass[reqno ,11pt]{amsart}
\usepackage[foot]{amsaddr}
\usepackage{graphicx}
\usepackage{bm}

\usepackage[usenames,dvipsnames]{xcolor}
\usepackage[paperwidth=7in,paperheight=10in,text={5in,8in},left=1in,top=1in,headheight=0.25in,headsep=0.4in,footskip=0.4in]{geometry}
\usepackage{subfigure}
\usepackage{lineno}
\usepackage{natbib} %this allows for styles in referencing
\usepackage{mathpazo}
\definecolor{linenocolor}{gray}{0.6}

\usepackage{fix-cm}

\usepackage{hanging}

\begin{document}

\title{Evolution of Cooperation Via Covert Signaling}
\author[Smaldino, Flamson, \& McElreath]{Paul E.~Smaldino$^{1,2}$, Thomas J.~Flamson$^1$, \and Richard McElreath$^{1,3}$}
\address{$^1$Department of Anthropology, University of California, Davis, Davis, CA, USA}
\address{$^2$Department of Political Science, University of California, Davis, Davis, CA, USA}
\address{$^3$Department of Human Behavior, Ecology and Culture, Max Planck Institute for Evolutionary Anthropology, Leipzig, Germany}
\email{paul.smaldino@gmail.com}

%\date{\today}

\maketitle

{\vspace{-6pt}\footnotesize\begin{center}Draft: \today\end{center}\vspace{12pt}}

%\linenumbers
%\modulolinenumbers[2]

%begin{abstract}{
\noindent \textbf{
Human sociality depends upon the benefits of mutual aid and extensive communication. However mutual aid is made difficult by the problems of coordinating diverse norms and preferences, and communication is harried by substantial ambiguity in meaning. Here we demonstrate that these two facts can work together to allow cooperation to develop, by the strategic use of deliberately ambiguous signals, covert signaling. Covert signaling is the transmission of information that is accurately received by its intended audience but obscured when perceived by others. Such signals may allow coordination and enhanced cooperation while also avoiding the alienation or hostile reactions of individuals with different preferences. Although the empirical literature has identified potential mechanisms of covert signaling, such as encryption in humor, there is to date no formal theory of its dynamics. We introduce a novel mathematical model to assess the conditions under which a covert signaling strategy will be favored. We show that covert signaling plausibly serves an important function in facilitating within-group cooperative assortment by allowing individuals to pair up with similar individuals when possible and to get along with dissimilar individuals when necessary.
This mechanism has broad implications for theories of signaling and cooperation, humor, social identity, and the evolution of human cultural complexity.}
%\end{abstract}

\vspace{12pt}


\newpage
\section*{Introduction}
Much of the research on human cooperation has focused on the free-rider problem: how to maintain cooperation when individuals' immediate interests are opposed to those actions that would maximize the total benefits to the group. However, in many cases individuals' interests are {\em aligned} rather than opposed, and these mutualistic scenarios may be equally important in understanding human social evolution \citep{skyrms2004stag, calcott2008other, tomasello2012two, smaldino2014cultural}. Even though there are no incentives for individuals to defect, mutualism is still a dilemma. When individuals {\em differ} in preferences, norms, or goals, the ability to efficiently coordinate breaks down. Therefore coordinating behavior, and in particular forming the reliable expectations of partner behavior that make coordination possible, is essential for the evolution of mutualism \citep{schelling1960strategy}.

Take, for example, the Battle of the Sexes game \citep{raiffa1957games}, in which there exist two equivalent Nash equilibria. Each player still prefers to coordinate rather than go it alone, but has a different idea of how to coordinate best. Each player would be better off finding another co-player with more aligned interests. Human societies are replete with dilemmas of this kind \citep{boyd1994norms}, and the need to coordinate extends to other forms of collective action as well \citep{ostrom2000norms}. Institutional mechanisms like punishment effectively convert other social dilemmas into coordination dilemmas, expanding their importance for understanding human sociality.

How can individuals assort on the basis of similarity in preferences, norms, goals, or strategies? Often these traits are difficult or impossible to directly observe. When preferences and norms are consciously held, individuals can merely signal their preferences. But often individuals are not conscious of their norms and preferences or realize their relevance too late to signal them.

One solution is the evolution of ethnic marking. Anthropologists have long recognized the importance of ethnic markers or tags in signaling group membership to improve cooperative outcomes \citep{barth1969}, and in recent years an extensive formal literature has developed exploring how these arbitrary signals can facilitate assortment on unconscious norms and preferences \citep{nettle1997, castro2007mutual, efferson2008coevolution, mace2005phylogenetic, mcelreath2003shared, moffett2013human}.

Language style and content can serve as a marker for social coordination \citep{nettle1997}. However it is striking that much communication is ambiguous. Is this ambiguity merely the result of constraints on the accuracy of communication? Here we propose that such ambiguity may
serve to facilitate coordination and thereby enhance cooperation within human societies. A basic problem with unambiguous signals is that they may foreclose less coordinated partnerships that may be of value in different contexts. In some situations, such as frequent or long-term endeavors, one is best served by engaging in homophilic assortment with a small set of similar partners who afford relatively more efficient coordination of behavior. In other contexts, different assortment outcomes may be desired, such as larger-scale cooperation for communal defense, differently-skilled partners for gains in trade, or when vying for the assistance of high-status individuals in political advancement. While overt signals of personal qualities like ethnic markers are useful in some contexts, where the adaptive problem is to delimit a set of partners who subscribe to the same broad behavioral norms and to categorically avoid interaction with those who do not, the ``all-or-nothing" character of such signals makes them inadequate for dealing with assortment {\em within} the groups delineated by these markers. Individuals will often benefit from not ``burning bridges" with less similar group members, so as to be able to draw on them for cooperation in other contexts. A signaling system that enables group members to communicate relative similarity only when similarity is high while retaining a shroud of ambiguity when similarity is low would facilitate assortment when the situation allows for it and still enable low-similarity assortment when the situation demands it.

Here we analyze the evolution of cooperative assortment by a form of signaling which satisfies these requirements and is known empirically to exist in nearly all human societies. Covert signaling is the transmission of information that is accurately received by its intended audience but obscured when perceived by others. It may na\"{i}vely appear that communication should have clarity as its goal. However, purposeful ambiguity is often strategic, allowing signalers both flexibility and plausible deniability \citep{eisenberg1984ambiguity, pinker2008logic}. Leaders may use ambiguous language to rally diverse followers under a common banner \citep{eisenberg1984ambiguity}, politicians may use vague platforms to avoid committing to specific policies \citep{aragones2000strategic}, and would-be suitors may mask their flirtations to be viewed innocuously if their affections are unreciprocated \citep{gersick2014covert}. What these discussions of ambiguity have in common is the assumption that all receivers will find the signals to be equally vague. In contrast, our discussion focuses on signals that will be clearer for some receivers and more ambiguous for others. A common example is ``dog-whistle politics'' \citep{lopze2014dogwhistle},
in which statements have one meaning for the public at large and a more specialized meaning for others. Such language attempts to transmit a coded message while alienating the fewest listeners possible.

A more precise and possibly much more common form of covert, within-group signaling is humor \citep{flamson2008encryption, flamson2013signals}. According to the encryption model of humor, a necessary component of humorous production is the presence of multiple, divergent understandings of speaker meaning, some of which are dependent on access to implicit information. Only those listeners who share access to this information can ``decrypt" the implicit understandings and understand the joke. Because the successful production of a joke requires access to that implicit information, humor behaves in manner similar to ``digital signatures" in computer cryptography, verifying the speaker's access to that information without explicitly stating it. By not explicitly declaring one's position within local variation, but instead signaling and assessing similarity on the basis of subtle and iterated cues that only like-minded group members can detect, individuals can engage in positive assortment in some contexts without burning bridges with potential allies in others.
While
not all humor necessarily has this form or function,
a substantial amount of spontaneous, natural humor does \citep{flamson2008encryption, flamson2013signals}.

Covert signaling can also facilitate assortment along dimensions of similarity more nuanced than discrete types or groups, which is a common restriction of ethnic markers or tags as they are typically discussed
\citep{mcelreath2003shared, antal2009evolution, cohen2013development, hammond2006evolution}. Jokes and other encrypted signals of identity can convey rich information about an individual's beliefs, goals, personality, proclivities, and history. Although any two individuals within a group should be able to cooperate when it is mutually beneficial to do so, pairs who are more similar along these trait dimensions should cooperate more effectively, generating larger benefits.

We propose that, by avoiding burned bridges, covert signaling serves an important function in facilitating cooperative assortment within groups by allowing individuals to pair up with similar individuals when possible and with dissimilar individuals when necessary. In the remainder of this paper, we precisely define the strategic logic of covert signaling in the context of opportunities for social assortment and coordination. We analyze the conditions for covert signaling to be preferred over overt signaling, in which information about an individual's traits is more transparent. Covert signaling is not always favored. For example, if it is possible to choose cooperative partners from a very large pool, overt signaling may be a more advantageous means of communication, as individuals will both avoid being paired with dissimilar partners and reap the added benefits that comes from {\em knowing} that a similarity exists \citep{chwe2001rational}. However, covert signaling often will be favored. It sacrifices maximal transparency for the sake of maintaining working relationships with dissimilar individuals. Covert signaling therefore may be an important part of a full explanation of both specific forms of communication and coordination, such as coded speech and humor, as well as the flexibility of human sociality more generally.

%%%%%%MODEL DESCRIPTION%%%%%%%%%%%%
\section*{Model Description}

We consider a large population of individuals who are generally cooperative toward one another. That is, we assume they have already solved the first-order cooperation problem of suppressing free riders, and can instead focus on maximizing the benefit generated by cooperation \citep{skyrms2004stag, calcott2008other, tomasello2012two, smaldino2014cultural}. Although individuals all belong to the same large group, they vary along many trait dimensions and thus share more in common with some individuals than others. Pairs of individuals whose trait profiles overlap to some threshold degree are deemed {\em similar} (S). Otherwise they are deemed {\em dissimilar} (--S). Pairs of similar individuals can more effectively coordinate, and so can obtain higher payoffs from cooperation. The probability that two randomly selected individuals will have similar trait profiles is given by $s$.

Individuals signal information about their trait profiles to the other members of the group for the purposes of facilitating future cooperative assortment. When a signal is successfully received, the recipient will update his disposition towards the sender, coming to {\em like} an individual whose signal indicates similarity and to {\em dislike} an individual whose signal indicates dissimilarity. When no signal is received, individuals maintain a neutral disposition toward the sender, reflecting uncertainty about their mutual similarity. Disposition is assumed to influence the effectiveness of assortment in addition to the influence of similarity: individuals who like each other can generate an even greater mutual benefit, while individuals who dislike each other are impaired.

Occasionally, individuals require assistance in a cooperative task, and attempt to form temporary cooperative partnerships to accomplish it. There are two distinct contexts: {\em free choice} and {\em forced choice} scenarios. In a {\em free choice} scenario, an individual has access to several members of her group from among whom she can choose a favored partner. In contrast, in a
{\em forced choice} scenario, an individual must seek help from whomever happens to be around.
Under these circumstances, it may be important not to have burned bridges, since this will limit the
likelihood of effective coordination.

The population dynamics consist of two stages: communication and cooperation.

\subsection*{Communication}
Each individual $i$ holds a {\em disposition} toward every other individual $j$. In the absence of any information about $j$'s trait profile, this disposition is neutral (N). If $i$ receives a signal from $j$ indicating similarity, this disposition will update to liking (L). If, on the other hand, $i$ receives a signal indicating dissimilarity, the disposition will update to disliking (D).

At the beginning of each generation, all individuals communicate information about their trait profiles to the members of the group. Individuals are either overt (O) or covert (C) signalers. Overt signalers attempt to be understood by all members of the population, and their signals are received by each individual with probability $r$. With probability $1-r$ a transmission error occurs, such as the receiver simply being out of the room when the signal was sent. Covert signalers share information about their trait profiles in such a way that it is obscured to dissimilar individuals and will only ever be received by similar individuals; they are therefore never actively disliked. However, covert signals are also more likely to be missed by their intended audience, such as when a joke falls flat, and so covert signals are received by similar individuals with probability $\gamma r$, where $\gamma$ represents the probability that a covert signal will be understood. After individuals signal about their trait profiles, all dispositions are updated in light of information received.

\subsection*{Cooperation}
Each individual is equally likely to seek out a cooperative partner at any given time, and so we can focus on the expected outcome for any particular cooperative scenario.
With probability $\delta$ the need for cooperation is ``dire" and the individual is forced to request cooperation from whomever is on hand. This will be an individual selected at random, as we assume high levels of mixing within the population. Otherwise, the individual can freely choose among a randomly sampled group of $m$ individuals, representing the subset of currently available partners (a more general case of the model is possible in which
two sample sizes are available for forced- and free-choice conditions, and is explored in the Appendix). An individual does not have direct access to information about her similarity to potential partners, but only to information about their disposition toward her. An individual has a strong preference for partners who like her, followed by individuals with a neutral disposition toward her, and only selecting a partner who dislikes her as a last resort.

There is no incentive to defect or refuse cooperation, and so individuals always grant requests for cooperation. However, the payoffs resulting from cooperative tasks are influenced by both similarity and disposition. There are four possible payoff outcomes. First, if the individual requesting assistance is disliked ($D$, which only occurs when the two individuals are dissimilar),
cooperation will be least coordinated, and the payoff will be minimal. We set this baseline payoff to zero without loss of generality. Second, if the disposition of the requested partner is neutral ($N$), the payoff depends entirely on similarity. We set the marginal benefit of {\em not} being disliked equal to one. This is the payoff received by dissimilar ($-S$) individuals when the requestee's disposition toward the requester is neutral ($N$). Third, the marginal benefit of being similar ($S$) in the same circumstance is given by $\alpha$. Note that this benefit does not have to present itself consciously; similar individuals may simply coordinate better by virtue of having more in common. Finally, being liked ($L$, which also implies similarity) generates an additional marginal benefit, $\beta$. This benefit might result from increased enthusiasm for the task given that each individual is working with someone they like, or because simply knowing about their similarity further increases their ability to coordinate \citep{chwe2001rational}. Formally, these payoffs are represented as follows:
\begin{align}
	&V(D) = 0, &
	&V(N,-S) = 1, \\
	&V(N, S) = 1 + \alpha, &
	&V(L) = 1 + \alpha + \beta. \nonumber
\end{align}

\subsection*{Fitness calculations}
Using the probability of each partnership characterized by the four possible combinations of similarity and disposition, we calculated the expected payoffs to overt and covert individuals. Our model is indifferent to whether these payoffs represent biological fitness or rather some welfare effect relevant to cultural replicator dynamics.

The nature of our model affords two particular mathematical conveniences in terms of payoff calculations. First, the probability of receiving a signal is independent of the receiver's own signaling strategy, and cooperative behavior is similarly independent of the requestee's signaling strategy. There are therefore no frequency-dependent effects. Second, requests for cooperation are always granted, so requestees obtain the same payoffs as requesters. We can therefore ignore payoffs to requestees in payoff calculations.

The fitness of each signaling strategy is a function of the seven model parameters (see Table 1).
The full derivation is given in the Appendix. The fitness of an overt signaler is
\begin{align}
W(O)		&= \delta \left[sr\beta + s \alpha + 1 - r(1-s) \right]  \\
 		& + (1-\delta) \left[\left(1 - (1-sr)^m \right)(1 + \alpha + \beta) \right.  \nonumber \\
 		& \left.  + \left((1-sr)^m - (r - sr)^m \right)
		(s\alpha + 1) \right],	\nonumber
\end{align}
and the fitness of a covert signaler is
\begin{align}
W(C) &= \delta \left[s\gamma r \beta + s \alpha + 1 \right] \\
	 	& + (1-\delta) \left[\left(1 - (1-s \gamma r)^m \right)(1 + \alpha + \beta) \right. \nonumber \\
		& \left.+ (1-s\gamma r)^{m-1}  \left(s(1-\gamma r)(1+\alpha) + 1 - s  \right) \right].\nonumber
\end{align}
Covert signaling is favored whenever $W(C) > W(O)$. This inequality can be rearranged with all parameters on the lefthand side and zero on the right. Covert agents will have a higher fitness than overt agents if and only if
\begin{align}
&	\delta r \left[ s\beta (\gamma - 1) + 1-s \right] \label{masterequation} \\
& \quad + (1 - \delta) (1 + \alpha + \beta) \left[ (1-sr)^m - (1 - s \gamma r)^m \right] \nonumber \\
& \quad + (1 - \delta)(1 - s \gamma r)^{m-1} \left[s(1 - \gamma r)(1+ \alpha) + 1 - s \right]\nonumber \\
& \quad - (1 - \delta) \left[(1 - sr)^m - (r - sr)^m \right](1 + s \alpha) > 0.\nonumber
\end{align}
The remainder of this paper is an explanation of what this inequality implies for the evolution of covert signaling. We take a two-pass approach to our analysis. First, we explore certain instructive cases for which intuitive analytical solutions are apparent. We follow this with numerical analyses to more fully characterize the model's behavior.

%\begin{table}[tp]
%\begin{center}
%\caption{Global model parameters.}
%\begin{tabular}{c l c}
%\hline
%Parameter 	&	 Definition	&	Range\\
%\hline
%$s$ 			&	Probability of similarity			& $[0, 1]$\\
%$\delta$ 			&	Rate of forced-choice scenarios			& $[0, 1]$\\
%$m$ 			&	Sample group size			& Positive integers	\\
%$r$ 			&	Baseline signaling efficacy			& $[0, 1]$ 	\\
%$\gamma$ 			&	Relative efficacy of covert signaling			& $[0, 1]$ 	\\
%$\alpha$ 			&	Marginal payoff of similarity			& $[0, \infty)$ 	\\
%$\beta$ 			&	Marginal payoff of being liked			& $[0, \infty)$ 	\\
%\hline
%\end{tabular}
%\end{center}
%\label{table_parameters}
%\end{table}%

%%%%%%%%%%%%%%%%% Results %%%%%%%%%%%%%%%%%%%%
\section*{Results}

Due to the structure of the payoff expressions, the dynamics of this system always lead to dominance by one signaling strategy or the other. One strategy or the other is favored at all frequencies. Therefore we seek conditions for covert signaling to dominate.

Covert signaling is favored by forced choice ($\delta>0$), the ability to reliably transmit covert signals ($\gamma > 0$), and when the population is sufficiently variable ($s<1$). These relationships are perhaps unsurprising, but the interactions among these factors can be complex, showing how even highly error-prone covert signals may evolve. In addition, the size of the pool of interaction partners, $m$, has a complex but ultimately intuitive relationship to the success of the covert strategy.

\subsection*{If the scenario is always forced-choice ($\delta = 1$)}
When choice is always forced, and assuming $r>0$ so that some signals of both types are received, covert signaling dominates when:
\begin{equation}
\frac{1 - s}{s} > \beta (1 - \gamma)
\label{eq:bg}
\end{equation}
The left side of the inequality is the odds of interacting with a dissimilar individual. The right side is the expected difference in marginal payoffs between overt and covert signaling when cooperating with a similar individual. Since covert signaling is prone to error $1-\gamma$ of the time, these errors reduce the range of conditions that favor covert signaling. If covert and overt signals are received with equal probability (i.e., $\gamma = 1$), then covert signaling will always be favored. However, errors can be very  common without displacing covert signaling, provided dissimilarity is sufficiently common and the benefit $\beta$ is not too large. We explore these relationships more thoroughly in the next section.

It should be noted that the situation when $\delta = 1$ is logically equivalent to one in which $\delta < 1$ but $m = 1$; that is, the number of individuals sampled under free choice conditions is one. This is also true algebraically. As choice becomes freer (i.e. $0 \le \delta <1$ and $m > 1$), it becomes possible for overt signalers to dominate under a larger range of conditions.

%%%%%%
\subsection*{If the sample group is very large ($m \gg 1$)}
The same relationship among these basic factors arises in very large groups.
In this case, whenever the situation is free choice, an individual will have a large group of possible partners from which to choose. The fitness of overt and covert individuals can then be approximated as follows:
\begin{align}
W(O)	&\approx \delta \left[sr\beta + s \alpha + 1 - r(1-s) \right]
 + (1-\delta) (1 + \alpha + \beta),\\	
W(C) &\approx \delta \left[s\gamma r \beta + s \alpha + 1 \right]
	 	+ (1-\delta)(1 + \alpha + \beta).
\end{align}
When groups are large, both strategies have equal fitness under free choice, because they can always find a similar partner who likes them. If $\delta > 0$, then the ability of covert signaling to dominate depends entirely on having a higher payoff under forced choice. The difference $W(C)-W(O)$ is:
\begin{equation}
\frac{1 - s}{s} > \beta (1 - \gamma)
\end{equation}
which is identical to Eq. \ref{eq:bg}.

\begin{figure}[tp]
\centerline{\includegraphics[width=.6\textwidth]{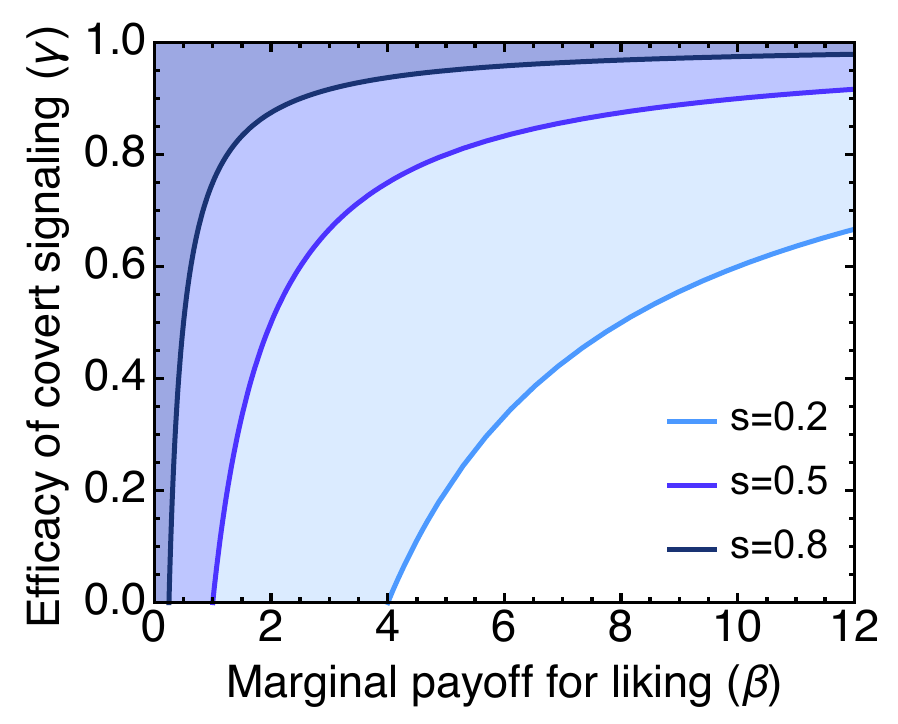}}
\caption{The minimal value of $\gamma$ needed for the evolution of covert signaling as a function of $\beta$ for several values of $s$. The shaded regions reflect values for which covert signaling dominates. This analysis assumes that $\delta > 0$ and $m$ is very large, so that free-choice scenarios always result in the maximal cooperative benefit for either signaling strategy.}
\label{fig_BetaVsGamma_s}
\end{figure}

It is an important result that covert signals can be highly noisy but still evolve. To demonstrate how noisy, we can re-write the condition for the dominance of covert signaling as the efficacy needed to overcome the benefit $\beta$ of being liked, for a given threshold of similarity $s$, as
\begin{equation}
\gamma > 1 - \left( \frac{1-s}{s} \right) \frac{1}{\beta}
\end{equation}
We plot this inequality in Figure \ref{fig_BetaVsGamma_s}. If $\gamma = 1$, so that covert signaling is as effective as overt signaling, then covert signaling will always dominate for large $m$ as long as $\delta > 0$. For imperfect covert signaling, noisier signals must be compensated by smaller benefits to being liked. If the benefit of being liked is sufficiently high, covert signaling cannot evolve. This relationship is also influenced by the threshold for similarity, $s$, or the probability that two randomly selected individuals will be sufficiently similar to generate enhanced coordination. When most individuals are similar, overt signaling is favored under a wide range of conditions, because there are many opportunities to reap the rewards of being liked and few opportunities to be disliked. Conversely, when similar individuals are rare, covert signaling is favored under a wider range of conditions, because pairings with dissimilar individuals will be common and it will therefore pay to avoid having burned bridges.

Restrictions on partner choice are commonplace, and the ability to choose a cooperative partner from among the entire population under normal circumstances is unrealistic. We therefore constrain the rest of our analysis to more arbitrary values of $m$.

%FIGURE 1 gamma vs. beta vs. s

\subsection*{Arbitrary group sizes}

\begin{figure}[tp]
\begin{center}
\centerline{\includegraphics[width=.95\textwidth]{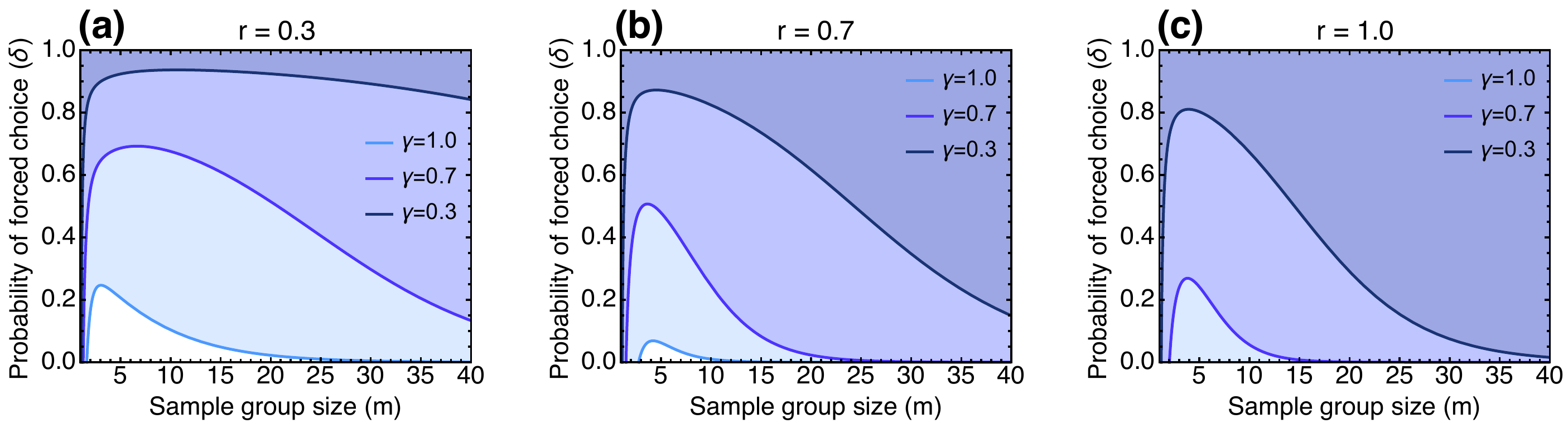}}
\caption{Triptych plot depicting critical values of $\delta$, above which covert signaling dominates, as a function of the sample group size, $m$. The shape of each curve varies with the values of the communication parameters, $r$ and $\gamma$.  Here, $s = 0.5$, $\alpha = 1$, $\beta = 1$.}
\label{fig_mVsDelta}
\end{center}
\end{figure}

As noted previously, a scenario where $m = 1$ is equivalent to $\delta = 1$, in which case covert signaling is strongly favored. As $m$ increases, the payoff difference between overt and covert signaling changes in a non-monotonic fashion, initially increasing sharply and then decreasing more gradually (Figure \ref{fig_mVsDelta}). What accounts for this non-monotonic relationship with group size?

%FIGURE 2 m vs. delta triptych

When $m = 1$, all interactions are with random individuals. As $m$ begins to increase, overt signalers gain an advantage, as they become better able to assort with similar individuals who like them. In free-choice scenarios, individuals can choose their interaction partners from a sample of size $m$. In these cases, overt signalers can more effectively sort similar from dissimilar individuals, making it both more likely that an individual can pair with a similar partner, and more likely that that partner will actively like the individual based on signaled common interests. This advantage is especially pronounced when overall communication rates are low and covert signals are highly error-prone.

\begin{figure}[tp]
\centerline{\includegraphics[width=.6\textwidth]{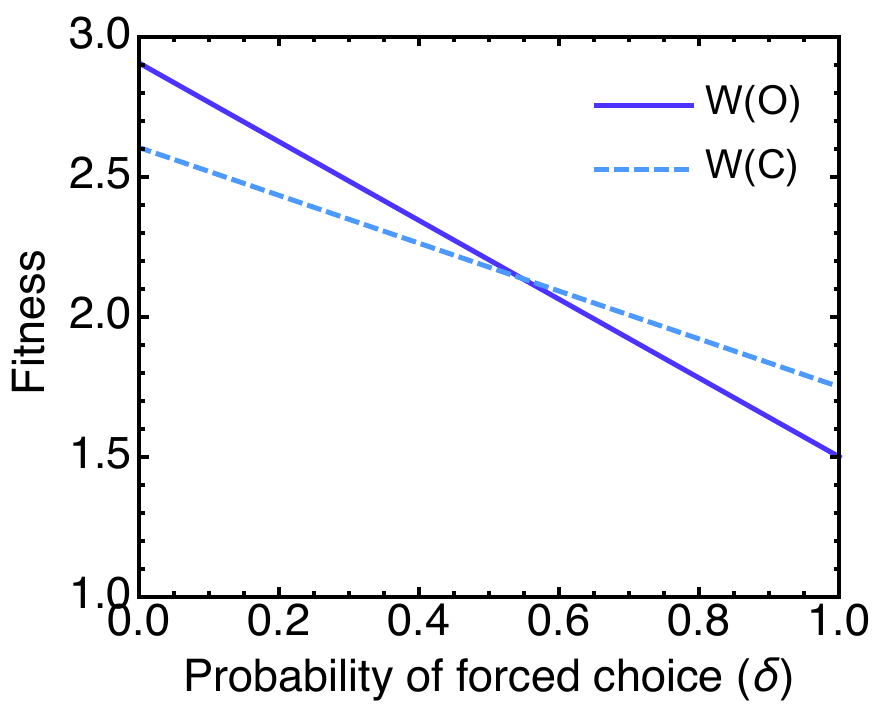}}
\caption{The fitness of overt signalers, W(O), and covert signalers, W(C), as a function of the probability of a
forced-choice scenario. This plot is merely illustrative, as the slope and intercepts of both lines depends on the remaining model parameters. Here, $s = 0.5$, $r = 1$, $\gamma = 0.5$, $m = 5$, $\alpha = 1$, $\beta = 1$.}
\label{fig_direfitness}
\end{figure}

In forced-choice scenarios, on the other hand, overt signalers risk being paired with a partner whom they have previously alienated, while covert signalers can avoid this fate because dissimilar individuals will remain neutral toward them.
In absolute terms, however, it is better for {\em everyone} when forced-choice scenarios are rare, because this increases the likelihood of an individual being paired with a similar partner who likes them. So, everyone's fitness decreases as forced-choice scenarios become more common, but the rate at which this occurs is lower for covert signalers, who are better equipped to successfully coordinate in these situations. Thus, as $\delta$ increases, there is often a critical value above which the fitness of covert signalers is greater than that of overt signalers (Figure \ref{fig_direfitness}). The precise value of this critical $\delta$ depends on the other model parameters.

%FIGURE 3 delta vs. fitness

\begin{figure}[h]
\centerline{\includegraphics[width=.6\textwidth]{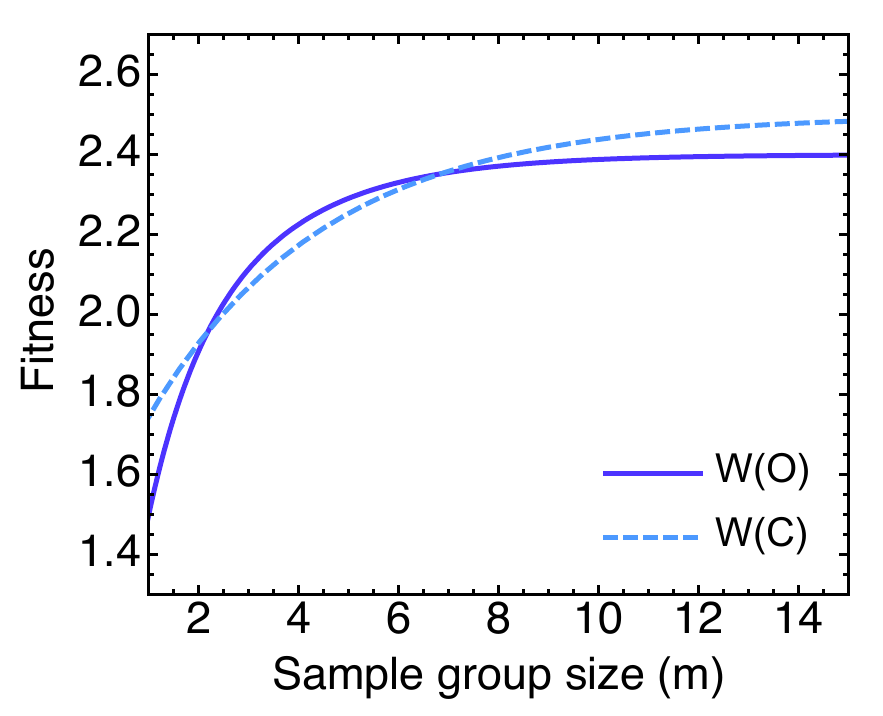}}
\caption{The fitness of overt signalers, W(O), and covert signalers, W(C), as a function of the sample group size, illustrating non-monotonicity of values supporting the evolution of covert signaling. Here, $s = 0.5$, $\delta = 0.4$, $r = 0.7$, $\gamma = 0.7$, $\alpha = 1$, $\beta = 1$.}
\label{fig_mfitness}
\end{figure}

As a consequence, as $m$ continues to increase, the marginal benefit to overt signalers' fitness declines as their payoffs reach an asymptote. Figure \ref{fig_mfitness}
illustrates that fitness for both strategies increases with $m$, but at different rates. For large enough $m$, both signaling strategies can effectively assort in free-choice situations, and the marginal advantage to covert signalers in forced-choice scenarios becomes the deciding factor.

%FIGURE 4 m vs. fitness

\begin{figure}[h!]
\centerline{\includegraphics[width=.6\textwidth]{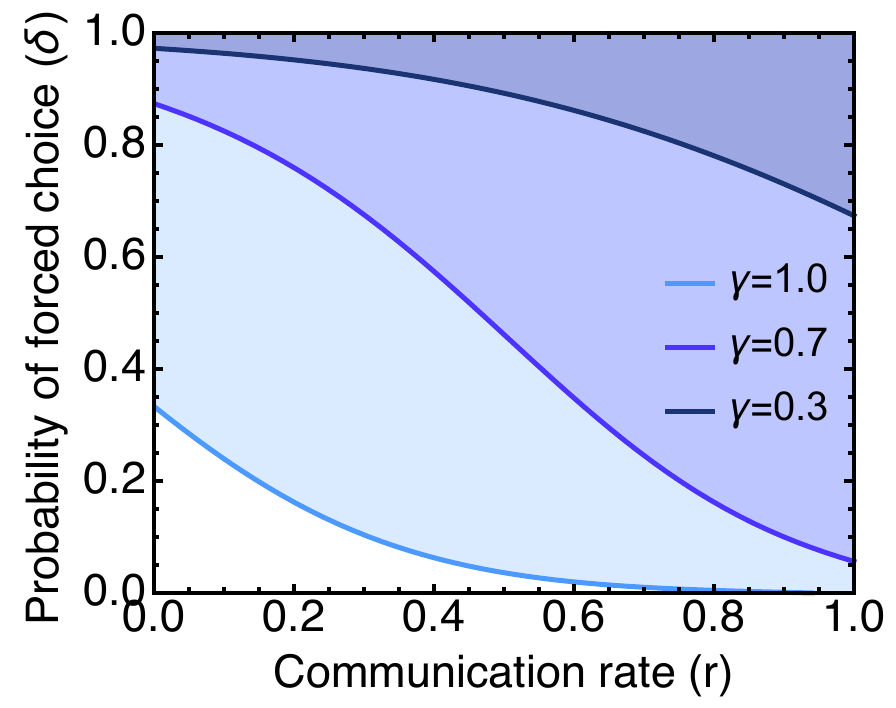}}
\caption{Critical values of $\delta$, above which covert signaling dominates, as a function of the communication rate, $r$, for several values of the efficacy of covert signaling, $\gamma$. Here, $s = 0.5$, $m = 10$, $\alpha = 1$, $\beta = 1$.}
\label{fig_rVsDelta}
\end{figure}

Figure \ref{fig_mVsDelta} shows the non-monotonic relationship of the threshold frequency of forced choice $\delta$ against group size $m$. The effectiveness of communication parameters $r$ and $\gamma$ are also varied, demonstrating their large influence on the evolution of covert signaling. Overt signaling is most favored when covert signaling is noisy ($\gamma \ll 1$) and communications rates are low ($r\ll 1$). This is more clearly shown in Figure \ref{fig_rVsDelta}, which plots $\delta$ against $r$ for three values of $\gamma$.

%FIGURE 5 r vs. delta

It is obvious that covert signalers should fare better when their signals are more likely to be received, i.e. when $\gamma$ is large. Less obvious is why overt signalers should have an advantage when communication rates are low across the board. When $r$ is small, information is rare, and hence access to information is at a premium. Overt signalers create higher-information environments, and as such are better at avoiding dissimilar partners and reaping the rewards of being liked by their cooperative partners. As signaling becomes more effective, both signaling strategies can manage to identify similar individuals and it becomes more important to avoid alienating the dissimilar individuals whom one might encounter. Covert signaling is thus favored in environments where it is easy to send signals to potential interaction partners.

\subsection*{No benefits to being similar or liked}

Covert signaling succeeds by not burning bridges with dissimilar cooperative partners. A stark demonstration of this is to note that covert signaling dominates when there are no marginal benefits to either similarity or liking, when $\alpha = \beta = 0$. In this case, the only difference in payoffs between signaling strategies stems from their ability to avoid being disliked. Here, covert signalers are at a clear advantage --- they are {\em never} disliked, and their expected fitness payoff is simply equal to one, the payoff from cooperating with someone who does not dislike them. Overt signalers, on the other hand, have payoffs that are diminished exactly by their probability of encountering a dissimilar individual who received their signal and therefore dislikes them:
\begin{align}
W(O) &= \delta \left[ 1 - r(1-s) \right] + (1 - \delta) \left[ 1 - [r (1 - s)]^m \right]
\end{align}
The above is always less than one. Covert signaling is always favored under these conditions.

%FIGURE 6 alpha vs. beta
\begin{figure}[tp]
\begin{center}
\centerline{\includegraphics[width=.95\textwidth]{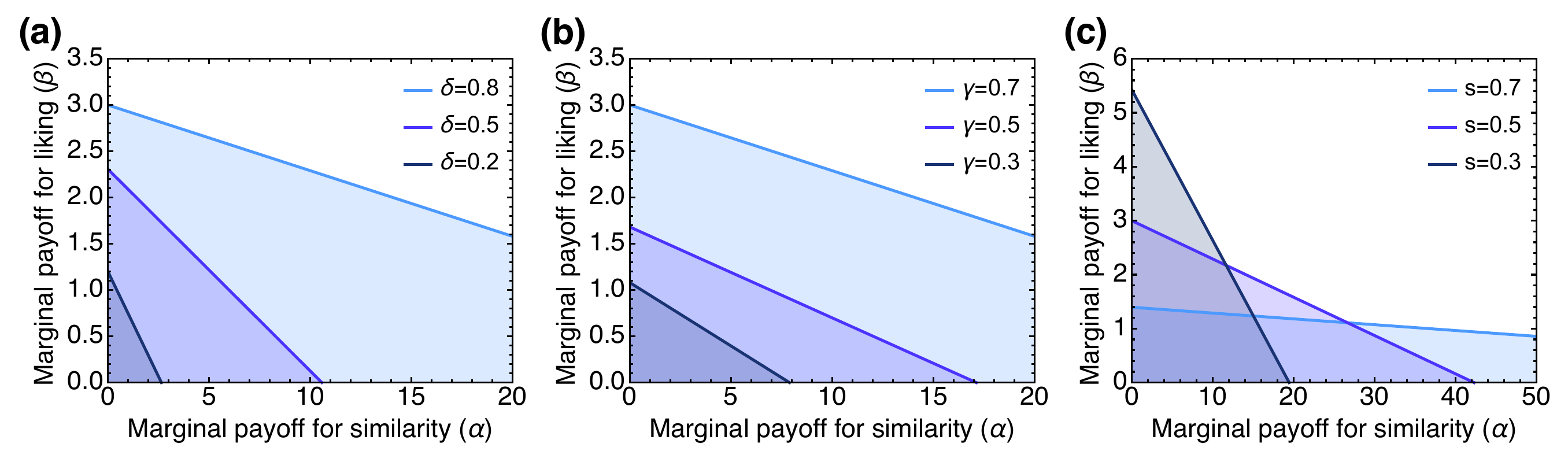}}
\caption{Critical values of $\alpha$ and $\beta$, below which covert signaling dominates and above which overt signaling dominates. For all plots, $m = 10$ and $r = 0.7$. (a) Plot for several values of $\delta$, with $\gamma = 0.7$ and $s=0.5$. (b) Plot for several values of $\gamma$, with $\delta = 0.8$ and $s=0.5$. (c) Plot for several values of $s$, with $\delta = 0.8$ and $\gamma = 0.7$. Please note the increased ranges on both axes.}
\label{fig_alphaVsBeta}
\end{center}
\end{figure}

If the benefits to being similar or (especially) being liked are sufficiently high, covert signaling cannot evolve (Figure \ref{fig_alphaVsBeta}), because overt signalers are better at assorting with similar individuals and communicating that similarity so that awareness can generate additional benefits. If the marginal payoff for being liked ($\beta$) is relatively small, then covert signaling can generally evolve for what are probably reasonable values for the marginal payoff for similarity, $\alpha$. Recall that in the population we are modeling, everyone is similar to {\em some} degree, and that the baseline value of neutral cooperation is one. Values of $\alpha$ many times greater than one may therefore be unlikely. The critical values of $\alpha$ and $\beta$ are highly dependent on the probability that a scenario will be
forced-choice, $\delta$, and on the efficacy of covert signaling, $\gamma$. If either of those latter parameters are low, covert signaling is favored for a narrower range of additional benefits. The critical values of $\alpha$ and $\beta$ also depend on the likelihood that random pairs of individuals are similar. When most individuals are similar, overt signaling dominates whenever the marginal benefit for being liked, $\beta$, is only slightly larger than the baseline benefit for not being disliked. In contrast, when similar individuals are rare, $\beta$ must be very large to overcome the benefits to covert signaling (Figure \ref{fig_alphaVsBeta}c).

%%%%%%%%%%%%%%%%% DISCUSSION %%%%%%%%%%%%%%%%%%%%
\section*{Discussion}
The dynamics of cooperation are more complicated than implied by models in which maximal benefits accrue to those who can simply avoid free riders. Not all cooperators are equal. Individuals vary, making assortment among cooperators important. Circumstances also vary. When individuals must occasionally collaborate with those outside their circles of friends, it can be critical to avoid burning bridges with dissimilar members of one's group. Covert signaling makes this possible, and this may be why it is sometimes observed in human societies, at both small and large scales.

We have shown that covert signaling is strongly favored when
average similarity is low, when
``dire" situations (in which partner choice is forced) are common, or when the marginal benefit to coordinating (being ``liked'') is small. We have also shown that covert signaling can be favored even when it is less effective than overt signaling at communicating similarity, because it simultaneously avoids communicating dissimilarity.

Our model also points to interesting transitions from inter- to intra-group assortment dynamics. As noted above, overt signaling systems are favored when the marginal benefit of being liked --- of working with similar others who are aware of that similarity, denoted here as $\beta$ --- is several times greater than the baseline benefit of cooperating with a dissimilar and neutral partner. This is precisely the kind of situation that is assumed to obtain in inter-group assortment, where overt signals such as ethnic markers are used to discriminate between similar and dissimilar individuals. In these between-group contexts, the difference between similar and dissimilar individuals is so great that attempting to coordinate with dissimilar others is not worth the effort, and one can afford to burn bridges with them in order to ensure that similar others are aware of their similarity \cite{mcelreath2003shared}. In fact, it might be argued that burning bridges with dissimilar out-group members is as much a goal of overt signals like ethnic markers as is attracting similar in-group members.

Intra-group assortment, however, is not simply a matter of scaling down inter-group dynamics. In this case, we must already presume some baseline level of similarity resulting from inter-group assortment; for there to be a group within which to assort, some degree of similarity should already be in place that defines that group, such as the shared interaction norms, communication systems, etc. 
that
ethnic markers are thought to ensure. The benefits of further assorting on the basis of more nuanced similarity are therefore likely to be marginal relative to random assortment within the group. In terms of our model parameters, very large $\alpha$ or $\beta$ favor overt signaling because they bring us out of the domain of intra-group assortment to that of inter-group assortment. With less dramatic marginal benefits, however, the costs of burning bridges with dissimilar group members make covert signaling worthwhile.

Relatedly, we emphasize that the probability of similarity, $s$, need not reflect some number of discrete types in the population, but can instead refer to a level of selectivity in how much a given pair of individuals needs to have in common in order to be considered ``similar.'' That is, $s$ refers to the proportion of the population that would be considered similar to a focal individual in a given context, with higher values indicating a looser concept of ``similar'' than lower ones (e.g., $s = 0.7$ indicates that the individuals are choosing partners on the basis of whatever criteria would include the 70\% of the group that are most similar to them). As Figure \ref{fig_alphaVsBeta}c illustrates, changes to $s$ can have a significant impact on the overall dynamics of the system, because $s$ essentially {\em defines} the concept of ``similarity." When $s = 1$, 100\% of individuals within the group will be considered ``similar," and the only useful signals will be overt ethnic markers that allow individuals to avoid dissimilar out-group members. As $s$ decreases, the criteria for considering a potential partner sufficiently similar to reap the benefits of enhanced coordination become stricter, as individuals deem a smaller proportion of the group worthy of homophilic assortment. The utility of covert signals increases as people become more choosy.

An interesting direction for future exploration is how these dynamics might respond to increased social complexity. In the larger and more complex societies associated with the development of agriculture, and particularly in the last few centuries, interactions with strangers have been increasingly frequent, necessitating strategies for temporary assortment \citep{johnson2000evolution, smaldinoInPressID}. In large, diverse populations, highly similar individuals should be rare, while the need for large scale cooperation in collective endeavors such as warfare, politics, or commerce would make it costly to burn bridges with these variously dissimilar partners. It is therefore likely that increases in social complexity would select for more complex covert signaling strategies.

For example, within complex industrialized societies, 
individuals often use Gestalt descriptions connoting a suite of information about the sort of person they are, which we call ``social identities." Identity signaling, whether through overt social markers or through more covert communication, can be used by individuals looking to find others similar to themselves and to avoid being mistaken for something they are not \citep{smaldinoInPressID, berger2008drives}. If the need to cooperate with dissimilar individuals is unlikely or if similar individuals are common, then overt declarations of identity should be expected. On the other hand, if burning bridges is both costly and likely given an overt signaling strategy, we should expect the relevant identity to be signaled much more subtly. There may be layers to how identity is signaled, with increasing levels of specificity signaled in increasingly covert ways, and without all received signals actively inducing a disposition of either liking or disliking toward the sender. A related signaling strategy, not covered by our model, might facilitate liking between similar individuals but only indifference otherwise. Using these ``semi-covert" signals, individuals would be aware of failures to match, but simply not care. Casual, coarse-grain identity signaling may often take this form, as in cases of fashion adoption or pop culture allegiances. It would be interesting to investigate how common these kind of semi-covert signals are in small-scale communities, as they seem pervasive in complex industrialized societies.

Ours is the first model of covert signaling. As such, it necessarily involves simplifying assumptions concerning the nature of signaling and cooperative assortment. For example, while we have allowed for covert signaling errors in the form of failed transmission to similar individuals, we have not included the converse form of error, where dissimilar individuals {\em are} able to detect the signal some of the time, and therefore update their disposition to disliking the covert signaler. Adding an additional parameter to account for this possibility does not qualitatively change our analysis. But it does create conditions where a non-signaling
``quiet" strategy could invade; see the Appendix for further discussion. In addition, we ignore the possibility of strategic action on the part of the receiver to either improve coordination or to avoid partnering with dissimilar individuals entirely. We assumed that a pairing of dissimilar partners would simply lead to an unsuccessful collaboration, but such a pairing might instead lead each individual to pursue more individualistic interests. At the population level, we assumed that all individuals had an equal probability of encountering similar individuals, and that all similar and dissimilar individuals were equivalent. In reality, some individuals may be more or less likely to encounter similar individuals, perhaps related to differences in the tendency to be conformity- versus distinctiveness-seeking \citep{smaldino2015social}, or reflecting minority-majority dynamics \citep{wimmer2013ethnic}.
Exploration of this variation opens the door to evaluating signaling and assortment strategies in stratified groups. All of these limitations provide avenues for future research that build upon the central findings reported here.

In a population where individuals vary and burning bridges is costly, overtly announcing precisely where one stands entails venturing into a zone of danger. Covert signaling, as in the case of humor or otherwise encrypted language, allows individuals to effectively assort when possible while avoiding burned bridges when the situation calls for partnerships of necessity.

%\section*{Acknowledgments}
%\noindent Thanks to Patrick Barclay, William Baum and John Bunce for helpful comments. This work was supported by National Science Foundation Grant BCS 1357240 (to T.F. and R.M.) and by the Division of Social Sciences Dean’s Office at the University of California Davis.

%---------------------------------------------------------------------------------------------------------------------------------------------------------------------------------------------------------------------
\newpage

\begin{center}
 {\bf
{\large APPENDIX\\~\large 
Covert Signaling Evolves By Avoiding Burned Bridges
}
}
\end{center}

%%%%%%%%%%%%%%%%% Fitness Calculations %%%%%%%%%%%%%%%%%%%%
\section{Fitness calculations}
Here we calculate the expected payoffs to overt and covert agents (see Table 1 for model parameters). First, some notation. We represent the probability of being liked by a randomly selected partner as $\Pr(L|O)$ for an overt signaler and $\Pr(L|C)$ for a covert signaler, with similar notation for disliking. The probability of an overt signaler randomly encountering a similar individual with a neutral disposition toward her will be written as $\Pr(N,S|O)$, and the probability of an overt agent randomly encountering a dissimilar individual with a neutral disposition toward her will be written as $\Pr(N,-S|O)$. Likewise for covert signalers.

\begin{table}[tp]
\begin{center}
\caption{Global model parameters.}
\begin{tabular}{c l c}
\hline
Parameter 	&	 Definition	&	Range\\
\hline
$s$ 			&	Probability of similarity			& $[0, 1]$\\
$\delta$ 			&	Rate of forced-choice scenarios			& $[0, 1]$\\
$m$ 			&	Sample group size			& Positive integers	\\
$r$ 			&	Baseline signaling efficacy			& $[0, 1]$ 	\\
$\gamma$ 			&	Relative efficacy of covert signaling			& $[0, 1]$ 	\\
$\alpha$ 			&	Marginal payoff of similarity			& $[0, \infty)$ 	\\
$\beta$ 			&	Marginal payoff of being liked			& $[0, \infty)$ 	\\
\hline
\end{tabular}
\end{center}
\label{table_parameters}
\end{table}%

For random interactions, such as those that occur under forced-choice conditions, the following are key probabilities:
\begin{align}
&	\Pr(L|O) = sr \\
&	\Pr(L|C) = s \gamma r\\
&	\Pr(N,S|O) = s(1-r)\\
&	\Pr(N,S|C) = s(1-\gamma r)\\
&	\Pr(N,-S|O) = (1-s)(1-r)\\
&	\Pr(N,-S|C) = (1-s)\\
&	\Pr(D|O) = (1-s)r\\
&	\Pr(D|C) = 0
\end{align}

For free-choice situations, we will need terms such as the probability of being liked by at least one agent in a set of $m$ randomly selected individuals. For an event which occurs with probability $q$, the general solution to this kind of problem -- i.e., the probability of at least one occurrence of the event in $m$ draws -- is equivalent to one minus the probability of zero draws: $1 - (1 - q)^m$.  Let $\Pr(L_1(m)|O)$ denote the probability that at least one individual in a set of $m$ likes the focal individual, given that the focal individual is an overt signaler. This works out to be:
\begin{align}
\Pr(L_1(m)|O) &= 1 - (1 - \Pr(L|O))^m
\end{align}
for overt signalers, and similar for covert signalers. For notational simplicity, we'll drop the $m$, so that $Pr(L_1|O) = Pr(L_1(m)|O)$.

To calculate the fitness of overt signalers, we need the following:
\begin{align}
&	\Pr(L_1|O) =  1 - (1-sr)^m\\
&	\Pr(N|O) = 1-r\\
&	\Pr(N_1|O) = 1 - r^m
\end{align}

We also need the joint probability that, given a sample group of $m$ individuals, there are zero individuals that like the focal individual {\em and} at least one individual who is neutral toward her. Let us denote the probability that there are zero individuals in a group of size $m$ that like the focal individual, given that she is an overt signaler, as $\Pr(L_0(m)|O)$. For overt signalers, this joint probability is given by the following, with the ``$(m)$" part omitted for convenience:
\begin{align}
\Pr(N_1, L_0 | O) 	&= \Pr(N_1 | L_0, O) \Pr(L_0|O)\\
		&= \left( 1 - \left[ 1 - \frac{\Pr(N|O)}{1 - \Pr(L|O)} \right]^m \right) (1 - \Pr(L|O))^m \nonumber \\
		&= \left( 1 - \left[1 - \frac{1 - r}{1-sr} \right]^m \right) (1 - sr)^m \nonumber \\
		&= (1 - sr)^m - [r(1-s)]^m	\nonumber								
\end{align}

The expected payoff for an overt signaler is therefore given by
\begin{align}
W(O) &= \delta \left[ \Pr(L|O)V(L) + \Pr(N,S|O)V(N,S) + \Pr(N,-S|O)V(N,-S) \right]\\
		&  \quad + (1-\delta) \left[ \Pr(L_1|O)V(L) \right. \nonumber \\
		 & \quad +  \left. \Pr(N_1, L_0|O) \left(\frac{\Pr(N,S|O)}{\Pr(N|O)}V(N,S) +\frac{\Pr(N,-S|O)}{\Pr(N|O)}V(N,-S) \right) \right] \nonumber \\
		&= \delta \left[sr(1+\beta) + s \alpha + (1-r) \right]
 + (1-\delta) \left[\left(1 - (1-sr)^m \right)(1 + \alpha + \beta) \right. \nonumber \\
 		& \left. \quad + \left((1-sr)^m - (r - sr)^m \right)
		\left(\frac{s(1-r)}{1-r}(1 + \alpha) + \frac{(1-s)(1-r)}{1-r}(1)   \right) \right], \nonumber
\end{align}
which reduces to
\begin{align}
W(O)		&= \delta \left[sr\beta + s \alpha + 1 - r(1-s) \right]
 + (1-\delta) \left[\left(1 - (1-sr)^m \right)(1 + \alpha + \beta) \right. \nonumber \\
 		& \left. \quad + \left((1-sr)^m - (r - sr)^m \right)
		(s\alpha + 1) \right].
\end{align}

%\vspace{-24pt}
We take the same approach for covert signalers, noting the added presence of the parameter $\gamma$, the efficacy of covert signaling. In addition to the probabilities for random interactions given above, we require the following probabilities for free-choice scenarios:
\begin{align}
&	\Pr(L_1|C) = 1 - (1 - s \gamma r)^m\\
&	\Pr(N|C) = 1 - s\gamma r\\
&	\Pr(N_1|C) = 1 - (s\gamma r)^m
\end{align}
We also require the joint probability that a covert signaler encounters zero individuals who like her and at least one neutrally disposed individual in a sample group of size $m$. We derive this here:
\begin{align}
\Pr(N_1, L_0 | C) 	&= \left( 1 - \left[ 1 - \frac{\Pr(N|C)}{1 - \Pr(L|C)} \right]^m \right) (1 - \Pr(L|C)^m\\
		&= \left( 1 - \left[1 - \frac{1 - s\gamma r}{1-s\gamma r} \right]^m \right) (1 - s\gamma r)^m \nonumber \\	
		&= (1 - s\gamma r)^m. \nonumber
\end{align}

The expected fitness for covert agents can now be calculated:
\begin{align}
W(C) &= \delta \left[ \Pr(L|C)V(L) + \Pr(N,S|C)V(N,S) + \Pr(N,-S|C)V(N,-S)  \right]\\
		& \quad + (1-\delta) \left[ \Pr(L_1|C)V(L) \right. \nonumber \\
		 & \quad + \left. \Pr(N_1, L_0|C) \left(\frac{\Pr(N,S|C)}{\Pr(N|C)}V(N,S) +\frac{\Pr(N,-S|C)}{\Pr(N|C)}V(N,-S) \right) \right]\nonumber \\
		 &= \delta \left[s\gamma r (1 + \alpha + \beta) + s(1 - \gamma r)(1+ \alpha) + 1 - s \right] \nonumber \\
		&\quad + (1-\delta) \left[\left(1 - (1-s\gamma r)^m \right)(1 + \alpha + \beta) \right.\nonumber \\
		&\quad \left.+ (1-s\gamma r)^m  \left(\frac{s(1-\gamma r)}{1-s\gamma r}(1 + \alpha) + \frac{(1-s)}{1-s\gamma r}(1)   \right) \right],\nonumber
\end{align}
which reduces to:
\begin{align}
W(C) &= \delta \left[s\gamma r \beta + s \alpha + 1 \right]
	 	+ (1-\delta) \left[\left(1 - (1-s \gamma r)^m \right)(1 + \alpha + \beta) \right.\nonumber \\ 
		&\quad \left.+ (1-s\gamma r)^{m-1}  \left(s(1-\gamma r)(1+\alpha) + 1 - s  \right) \right].		
\end{align}

Covert signaling will be favored whenever $W(C) > W(O)$, or when $W(C) - W(O) > 0$. Expanding out this inequality, we derive the following condition for the evolution of covert signaling:
\begin{align}
&	\delta r \left[ s\beta (\gamma - 1) + 1-s \right] \\
& \quad + (1 - \delta) (1 + \alpha + \beta) \left[ (1-sr)^m - (1 - s \gamma r)^m \right] \nonumber \\
& \quad + (1 - \delta)(1 - s \gamma r)^{m-1} \left[s(1 - \gamma r)(1+ \alpha) + 1 - s \right]\nonumber \\
& \quad - (1 - \delta) \left[(1 - sr)^m - (r - sr)^m \right](1 + s \alpha) > 0.\nonumber
\end{align}

%\newpage
%%%%%%%%%%%%%%%%% A more general model %%%%%%%%%%%%%%%%%%%%
\section{Generalization to two sampling scenarios}
In the model presented in the main text, we contrasted two possible scenarios in which individuals required assistance in a cooperative task. Under free-choice conditions, individuals were exposed to a sample group of $m$ individuals. Although the focal individual could not access information about similarity directly, she could ascertain the sample individuals' dispositions toward her. On the other hand, in forced-choice scenarios, the focal individual was forced to cooperate with a randomly selected individual.

A more general version of this model is one in which under more favorable conditions, a focal individual can select a cooperative partner from a sample group of $m_1$ individuals, while under less favorable conditions, she has access to a smaller sample group of size $m_2$, such that $m_2 \le m_1$. In this case, the expected fitness for overt signalers is given by
\begin{align}
W(O)		&= \delta  \left[\left(1 - (1-sr)^{m_2} \right)(1 + \alpha + \beta) \right.\\
 		  	&\quad \left.+ \left((1-sr)^{m_2} - (r - sr)^{m_2} \right)(s\alpha + 1) \right]\nonumber \\
			&\quad + (1-\delta) \left[\left(1 - (1-sr)^{m_1} \right)(1 + \alpha + \beta) \right.\nonumber \\
 		  	&\quad \left. + \left((1-sr)^{m_1} - (r - sr)^{m_1} \right)(s\alpha + 1) \right].\nonumber
\end{align}
The expected fitness for covert signalers is similarly given by
\begin{align}
W(C) &= \delta \left[\left(1 - (1-s \gamma r)^{m_2} \right)(1 + \alpha + \beta) \right.\\
		&\quad \left.+ (1-s\gamma r)^{m_2-1}  \left(s(1-\gamma r)(1+\alpha) + 1 - s  \right) \right] \nonumber \\
	 	&\quad + (1-\delta) \left[\left(1 - (1-s \gamma r)^{m_1} \right)(1 + \alpha + \beta)\right.\nonumber \\
		&\quad \left. + (1-s\gamma r)^{m_1-1}  \left(s(1-\gamma r)(1+\alpha) + 1 - s  \right) \right].\nonumber
\end{align}
The model in the main text is recovered when $m_2 = 1$.

%\newpage
%%%%%%%%%%%%%%%%% A more general model %%%%%%%%%%%%%%%%%%%%
\section{Quiet: Comparison to a strategy of non-signaling}

Consider a type of strategic agent in our model that communicates no information about herself to others. She will never be disliked, nor will she be liked. This is equivalent to imagining a covert signaler whose signaling efficacy is $\gamma = 0$. From the equation for the fitness of covert agents, it is easy to show that the expected payoff to these ``quiet" (Q) agents is simply
\begin{equation}
W(Q) = 1 + s \alpha.
\end{equation}
Because quiet agents communicate no information, they interact with similar partners only by chance, with a probability equal to the prevalence of similar individuals in the population.

For our model, it is clear that, as long as $\gamma > 0$, covert agents will outperform quiet agents --- i.e., $W(C) > W(Q)$ --- because covert agents similarly avoid being disliked but are better able to assort with similar partners  and occasionally reap the rewards of being liked. The quiet strategy therefore represents a lower bound to the fitness of covert signaling. As our intention is to explore the conditions under which covert signaling is favored, we therefore do not investigate the quiet strategy in further detail here.

In the model presented in this paper, we assumed that all transmission errors were ones of omission; that is, those in which a signal failed to either transmit or be received. However, one can also imagine situations in which covert signals are occasionally detected by {\em dissimilar} individuals, such as when someone 
understands a derogatory joke, but does not share the speaker's stance towards the targets. This could result in covert signalers suffering the same cost of being disliked as overt signalers, and creating 
conditions under which it would be preferable to remain silent rather than attempting to signal covertly. An analysis of such a model, in which overt, covert, and quiet strategies could all compete, is beyond the scope of the present paper, but is a target for research from our group in the near future.

%\clearpage
\bibliographystyle{newapa}
\bibliography{covert}

\end{document}